# High-precision optical fiber sensing beyond 1000°C

**Mohan Wang,[1] Patrick S. Salter,[1] Frank P. Payne,[1] Adrian Shipley,[2] Igor N. Dyson,[1] Tongyu Lui,[1] Tao Wang,[3,4] Kaihui Zhang,[3] Jian Zhang,[3] Zhitai Jia,[3] Stephen M. Morris,[1] Martin J. Booth,[1] and Julian A. J. Fells[1,*]**

[1]*Department of Engineering Science, University of Oxford, Parks Road, Oxford, OX1 3PJ, UK*
[2]*Rolls-Royce Plc, Derwent Building, 5000 Solihull Parkway, Birmingham Business Park, Birmingham, B37 7YP, UK*
[3]*State Key Laboratory of Crystal Materials, Shandong University, 27 South Shanda Road, Jinan, Shandong, 250100, China*
[4]*Jiangsu Jingying Optoelectronics Technology Co. LTD, 10 East Zhujiang Road, Xuzhou, Jiangsu, 221116, China*
*\*Corresponding author: julian.fells@eng.ox.ac.uk*

**Sapphire fiber can withstand around 2000°C, but it is multimoded, giving poor precision sensors. We demonstrate a single-mode sapphire fiber Bragg grating temperature sensor operating up to 1200°C. The repeatability above 1000°C is within ±0.08%.**

Optical fiber sensors are commonly used for remotely monitoring a wide range of infrastructure and machinery. High-quality monitoring of environmental parameter changes can improve safety, extend lifetime, lower environmental impact, and prevent potential failures at an early stage. However, there are many applications, particularly in the energy, aerospace and space sectors, where operating temperatures are beyond the 1000°C limit of silica fiber sensors [1].

An important class of sensor is the fiber Bragg grating (FBG), which is a periodic modification of the effective refractive index along the length of an optical fiber. An FBG reflects light predominantly at a wavelength determined by the product of the effective refractive index and the physical period. As these properties are temperature dependent, FBGs can operate as temperature sensors. However, as temperatures approach 1000°C, there is considerable drift in the wavelength of reflection in silica fibers over time [1]. As an alternative, sapphire optical fiber can withstand far higher temperatures up to 2000°C. Sapphire FBGs have been demonstrated up to 1900°C [2], with long-term stability tests up to 1400°C [3]. Sapphire fiber is a high refractive index single-crystal without a cladding, and therefore highly multimoded. Since different modes have different effective indices, Bragg gratings in sapphire fiber typically have a very broad reflection spectrum [4]. This severely compromises the measurement precision attainable and reduces the number of sensors which may be multiplexed. Furthermore, changes in temperature and vibration result in fluctuation in the spectral power distribution. There have been attempts to create a cladding around a sapphire fiber by adding additional coatings [5] and also creating an inner cladding within the fiber by irradiation [6]. Another approach has been to reduce the core diameter by etching in hot acid [7], However, in all these cases, the resulting fiber was still multimode. There have been numerical studies on suppressing higher order modes in sapphire fiber by having holes in the side of the fiber [8], but these devices have not been realized experimentally.

We previously demonstrated a 1 cm single-mode sapphire FBG [9], but longer devices are required for sensing. Subsequently, a 2 mm single-mode helical Bragg grating was reported in multimode sapphire fiber [10] and very recently silica single-mode fiber (SMF) was spliced to launch into the fundamental mode of multimode sapphire fiber [11]. However, although it may be possible to weakly launch into a particular mode of multimode fiber, coupling to other modes will occur and vary with environmental conditions [12].

In this Letter, we use high quality sapphire fiber growth and an improved laser-writing process to extend the length of the single-mode sapphire fiber. This has allowed sensors in fibers of sufficient length to be tested at ultra-high temperatures.

The sensor design is schematically illustrated in Fig. 1. The sapphire fiber is 4 cm long with a single mode waveguide along its entire length. As femtosecond laser exposure results in a reduction in the sapphire refractive index, this waveguide is a depressed cladding waveguide. The waveguide is formed by leaving the core unmodified and laser-writing the material around the core to reduce its refractive index, thereby forming the cladding. There is a 1 cm long Bragg grating at one end of the fiber within the core. The opposite end of the sapphire fiber is spliced to standard silica SMF.

We grew 100 μm diameter sapphire fiber using a custom laser-heated pedestal growth system [13]. The c-axis is along the fiber length. The fiber was processed using a regenerative femtosecond laser system (Pharos SP-06-1000-PP) delivering

170 fs pulses at 515 nm. The output beam was expanded using a telescope and projected onto a spatial light modulator (SLM). A pre-distorted phase pattern was applied to the SLM to correct for aberrations inside the laser beam delivery system. The pre-distorted beam was projected onto the pupil plane of the objective through a 4-f imaging telescope. The linearly polarized laser beam was focused onto the fiber through a ×63 1.25 NA oil objective through immersion oil (Cargille Labs, $n$=1.7450). A multilayer depressed cladding single-mode waveguide was fabricated based on a refinement of our previous design [9], using a pulse energy of 28 nJ.

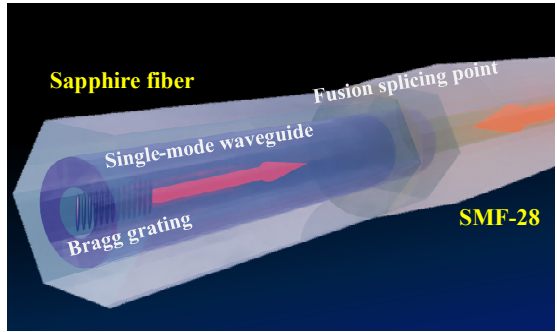

Fig. 1. Schematic of the sapphire sensor.

The design consisted of 215 tracks with the grating axially in the center. The structure was fabricated using a three-step process. First, the bottom layers of the waveguide were written at 2 mm/s and a repetition rate of 100 kHz. Then, a second-order point-by-point Bragg grating was embedded within the center of the waveguide, spanning the first 1 cm of the 4 cm single-mode waveguide. Modulated bursts utilizing the stage's pulse synchronization output were used to create a designed grating period of 887.64 nm with 22 pulses per pitch. Finally, the top layers of the waveguide were written (with the same parameters as the bottom layers) to enclose the grating. After fabrication, the fiber was polished on both ends.

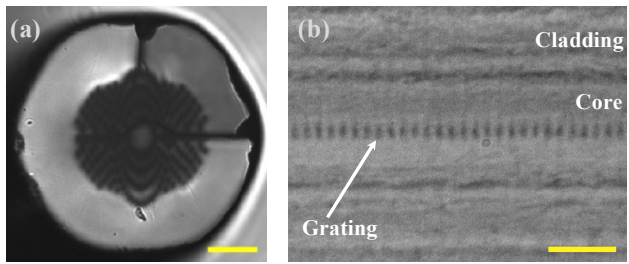

Fig. 2. (a) End image of the waveguide, the scale bar is 20 μm and (b) top view of the waveguide and grating, the scale bar is 5 μm.

Figure 2 shows microscope images of the single-mode waveguide. The view from the fiber end-face is shown in Fig. 2(a). Two cracks were incurred by stress introduced during polishing and are confined to within close proximity of the fiber end facets. Through tuning the waveguide design and fabrication conditions, no cracks were observed in or around the Bragg grating and waveguide region along the 4 cm sensor, except for around the facets. A side-view is shown in Fig. 2(b). The boundaries of the core and cladding regions can be seen, with the point-by-point grating clearly visible in the center of the core.

Standard silica SMF (SMF28e+ equivalent) was butt-coupled to the single-mode sapphire fiber to measure the mode profile. The near-field mode profile at 1550 nm from the sapphire fiber waveguide is shown in Fig. 3(a). The mode field diameter (MFD) was calculated from a Gaussian fit to obtain the full width at $1/e^2$ intensity. The MFD was between 8.5 and 8.7 μm, which is sufficiently close to the 10.4±0.5 μm MFD of standard SMF for reasonable coupling efficiency. For reflection and transmission measurements, a swept tunable laser (Agilent 8164A) was coupled to an optical circulator to inject light into the grating, with the reflected light monitored on a photodetector within the 8164A. A manual polarization controller was used to adjust the laser polarization.

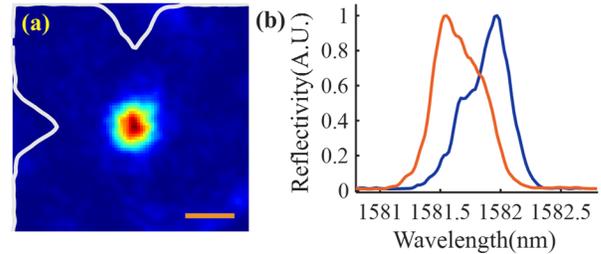

Fig. 3. (a) Transmission mode field, the scale bar is 10 μm and (b) the spectra of the two orthogonal polarizations measured at 1200°C.

Although multimode sapphire fibers have previously been spliced to multimode silica fiber [14], splicing the single-mode sapphire fiber to silica SMF is much more difficult due to the need for matched mode fields and precise core alignment. As splicer imaging systems do not have aberration compensation for the high index sapphire, it is not possible to use visual core alignment. The use of active alignment by monitoring transmitted light was prevented by the large Fabry-Pérot ripple observed when there is an air-gap between the two fibers. Instead, we developed a technique whereby the Bragg reflection spectrum was monitored from the silica SMF.

The sapphire fiber end furthest from the grating was spliced to standard silica SMF using a fusion splicer (Sumitomo Type-72C) in manual mode. The sapphire fiber and the silica SMF were placed in the fiber chucks with a small gap between and were initially aligned by eye using the motorized positioners. The Bragg reflection spectrum was monitored using the reflection measurement setup to inject light into the grating via the silica SMF. Using the splicer positioning stages, precise core alignment was achieved through an iterative manual fiber adjustment process, by maximizing the magnitude of the reflected peak in wavelength. Once the optimum fiber alignment was obtained, the arc fusion process was initiated, with an arc time of 2 seconds and a pre-fuse time of 0.15 seconds.

The sapphire fiber sensor with its lead-in silica SMF tail was inserted through a ceramic tube opening into a box furnace with a maximum temperature of 1200°C. A Type-K thermocouple was placed in proximity to the sensor. Temperature cycling tests were then performed over a period

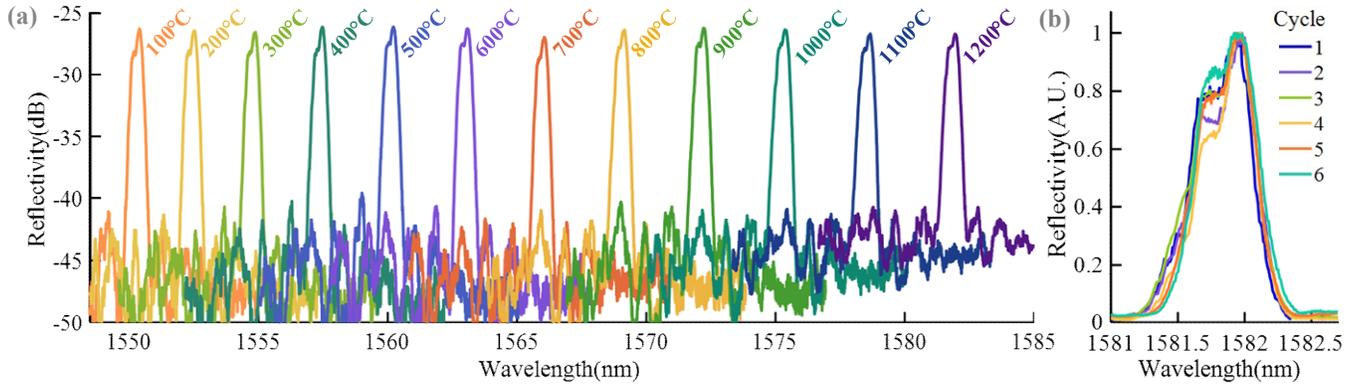

Fig. 4 (a) Spectra of the sapphire FBG during one heating cycle and (b) spectra at 1200°C for repeated heating cycles with normalized magnitude. See Datafile 1 for the underlying data.

of 9 days. The sensor was initially taken to 1000°C and back down. It was then taken through six repeated heating and cooling cycles (Cycles 1-6) in 100°C steps to 1200°C. The temperature was held at 1200°C for 1 hour during Cycle 1 for annealing. At each temperature, the reflection spectrum was measured over a 10 nm range with a 10 pm step size.

The input polarization was adjusted to reveal two orthogonal polarization modes. Figure 3(b) shows the two polarization modes at 1200°C, at a resolution of 20 pm. The peaks are separated by 0.43 nm. The full-width half-maximum (FWHM) bandwidths of the two polarization modes were measured to be 0.50 nm and 0.55 nm, respectively. The existence of the birefringence is likely a result of the anisotropic refractive index distribution of the Bragg grating. Commercial interrogators typically have a polarization scrambler. As the modes are distinct, it would be possible to determine the wavelength of both peaks with a scrambled input polarization and take a mean value. The slow axis was used for the subsequent measurements.

Figure 4(a) shows spectra at each temperature overlaid for Cycle 5. There is an increase in the noise background as the temperature increases, but there is a signal-to-noise ratio (SNR) of over 16 dB across the full range. Figure 4(b) shows the overlapped spectra at a temperature of 1200°C for all cycles, showing a consistent peak wavelength, with only slight fluctuation, recorded at a resolution of 10 pm.

Figure 5 shows the Bragg wavelength against temperature, for Cycles 1-6. We used a quadratic fit $\lambda_B = A + BT + CT^2$, with $A$, $B$, and $C$ found to be 1548.09 nm, 21.60 pm/°C, and 0.0056 pm/°C2 respectively, which are consistent with the literature [15]. The sensitivity varies between 22.7 to 35.1 pm/°C Also shown in Fig. 5 is the residual temperature deviation at each temperature point, obtained by subtracting the mean Bragg wavelength and dividing it by the gradient of the characteristic curve at that point. The variation was within ±2.5°C. Much of this variation will be due to the furnace temperature varying over the 15 minute time interval it took to take each spectrum. The furnace did not have cooling, hence the control system had much better stability at higher temperatures. There is also the ±0.5°C accuracy of the thermocouple to consider. However, above 1000°C, the variation is within ±0.8°C, or within ±0.08%, which is significantly better than previous measurements [4]. The fiber length is around that needed to access a gas turbine and the high SNR gives potential to increase the length.

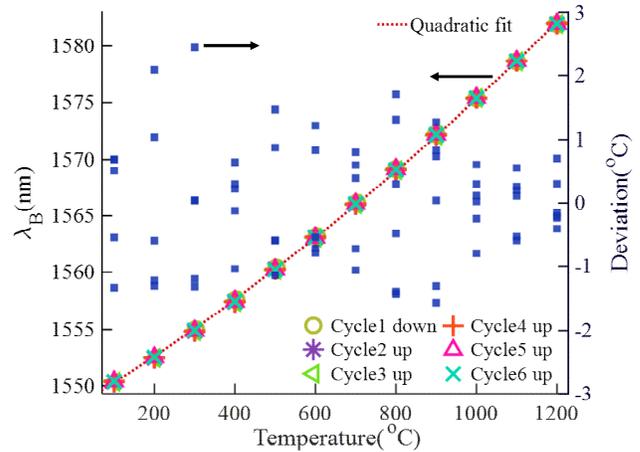

Fig. 5. Left axis: Bragg wavelength against temperature for 6 heating and cooling cycles, with a quadratic fit calibration curve.; Right axis: Temperature measurement deviation from the mean reading using the calibration curve. See Datafile 2 for the underlying data.

To conclude, a 4-cm long single-mode sapphire fiber Bragg grating temperature sensor with a standard single-mode silica fiber tail was fabricated. The sensor was tested in a furnace through 6 temperature cycles up to 1200°C, showing a sensitivity between 22.7 to 35.1 pm/°C. The repeatability above 1000°C was within ±0.08%. This demonstration shows the potential for single-mode sapphire fiber sensors for ultra-extreme environments.

**Funding.** Engineering and Physical Sciences Research Council (EP/T00326X/1).

**Disclosures.** The authors declare no conflicts of interest.

**Data availability.** Data underlying the results presented in this Letter are available in Datafile 1 and Datafile 2 of the supplementary information.